\documentclass[onecolumn,amsmath,amssymb,11pt,superscriptaddress,nofootinbib]{revtex4}

\usepackage{graphicx}
\usepackage{amsmath,amssymb,amsfonts,amsthm,stmaryrd,mathtools,bm,physics,tensor}
\usepackage{color}
\usepackage{tikz}
\allowdisplaybreaks[1]

\usepackage[bookmarks,linktocpage, colorlinks=true, plainpages = false, citecolor = treegreen,  linkcolor=darkblue, urlcolor = darkblue, filecolor = blue]{hyperref} 

\def\be#1\ee{\begin{align}#1\end{align}}

\def\ba{\begin{eqnarray}}
	\def\ea{\end{eqnarray}}

\definecolor{tealgreen}{rgb}{0.0, 0.5, 1.0}
\definecolor{darkblue}{rgb}{0., 0.4, 0.8}
\definecolor{cadmiumred}{rgb}{1., 0., 0.22}
\definecolor{treegreen}{rgb}{0., 0.7, 0.3}

\newcommand{\noteJB}[1]{{\textcolor{darkblue}{#1}}}

\begin{document}
	
\title{
From area metric backgrounds to the cosmological constant \\ 
and corrections to the Polyakov action\\
}

\author{Johanna Borissova}
\email{jborissova@pitp.ca}
\affiliation{Perimeter Institute for Theoretical Physics, 31 Caroline Street North, Waterloo, ON, N2L 2Y5, Canada}
\affiliation{Department of Physics and Astronomy, University of Waterloo, 200 University Avenue West, Waterloo, ON, N2L 3G1, Canada}

\author{Pei-Ming Ho}
\email{pmho@phys.ntu.edu.tw}
\affiliation{Department of Physics and Center for Theoretical Physics, \\
National Taiwan University, Taipei 106, Taiwan,
R.O.C.} 
\affiliation{Physics Division, National Center for Theoretical Sciences, \\
Taipei 106, Taiwan,
R.O.C. }

\begin{abstract}
	
Area metrics and area metric backgrounds provide a unified framework for quantum gravity. They encode physical degrees of freedom beyond those of a metric. These non-metric degrees of freedom must be suppressed by a potential at sufficiently high energy scales to ensure that in the infrared regime classical gravity is recovered. On this basis, we first study necessary and sufficient algebraic conditions for an area metric to be induced by a metric. Second, we consider candidate potentials for the area metric and point out a possible connection between the reduction of area metric geometry to metric geometry on the one hand, and the smallness of the cosmological constant on the other. Finally, we consider modifications of the Nambu-Goto action for a string, from a metric background to an area metric background. We demonstrate that area metric perturbations introduce an interaction corresponding to a singular vertex operator in the classically equivalent Polyakov action. The implications of these types of vertex operators for the quantum theory remain to be understood.
	
\end{abstract}

\maketitle
\tableofcontents

\section{Introduction}

Area metric geometry~\cite{Schuller:2005yt,Schuller:2005ru,Punzi:2006hy,Punzi:2006nx,Ho:2015cza,Borissova:2023yxs} is a generalized approach to spacetime geometry traditionally described in terms of a length metric. Whereas a length metric is represented by a rank-two tensor, which measures the norms of vectors and angles between two vectors, an area metric is represented by a rank-four tensor, which measures the norms of bivectors and dihedral angles between two bivectors. A given length metric always gives rise to an induced area metric. However, in dimensions $d>3$, a generic area metric exhibits more degrees of freedom than an ordinary metric and thereby defines a much richer structure to describe background spacetime than a metric.

The relevance of area metrics extends across various areas of physics and, in particular, approaches to quantum gravity. The Nambu-Goto action for a free string, as well as the action for the electromagnetic field and the Yang-Mills action, are naturally described by an antisymmetric product of metrics that defines an induced area metric. Taking this area metric to be generic and not induced by a metric, defines these actions on an area metric background, on which area connections and curvature tensors can be introduced~\cite{Schuller:2005ru,Schuller:2005yt,Ho:2015cza}. 
In particular, in string theory, at least from the viewpoint of the Nambu-Goto action or the Schild action, only the notion of area, not the notion of length, is needed for the worldsheet theory~\cite{Ho:2015cza}. The area metric is a more natural geometric notion for strings, while the metric is for particles.
Moreover, area metrics also appear in holography and the reconstruction of geometry from entanglement~\cite{Ryu:2006bv,Swingle:2017blx}.

Notably, area variables provide the fundamental variables in loop quantum gravity~\cite{Rovelli:1997yv} and spin-foam path integrals for quantum geometry~\cite{Perez:2012wv}. Therein, four-dimensional area metrics appear at the microscopic level in the twisted geometry of a coherent four-simplex~\cite{Dittrich:2023ava}, as well as macroscopically in the continuum limit of the Area-Regge action~\cite{Rovelli:1993kc,Barrett:1997tx} which describes the semiclassical regime of effective spin foams~\cite{Asante:2020qpa}, see~\cite{Dittrich:2021kzs,Dittrich:2022yoo}. In the latter context, the additional degrees of freedom of the area metric, beyond those of an ordinary metric, can be understood as arising due to a second-class quantum algebra of constraints~\cite{Dittrich:2008ar,Dittrich:2010ey,Dittrich:2012rj}. The presence of second-class constraints requires these to be imposed only weakly and thereby leads to an enlarged configuration space over which the gravitational spin foam path integral is defined. In~\cite{Borissova:2022clg} it has been shown that modified non-chiral Plebanski theories~\cite{DePietri:1998hnx,Alexandrov:2008fs,Speziale:2010cf}, in which a subset of the simplicity constraints is replaced by a potential, can be understood as a classical analog of the mechanism of weak imposition of constraints at the quantum level. In particular, modified non-chiral Plebanski theories were seen to reproduce the results found in the continuum limit of the Area-Regge action~\cite{Dittrich:2021kzs,Dittrich:2022yoo}, and to provide a natural non-perturbative framework for area metrics and area metric actions, see~\cite{Borissova:2022clg}. 

In a perturbative analysis, area metric actions to second order in fluctuations and derivatives have been studied in~\cite{Borissova:2023yxs}. The effective actions for the subset of area metric degrees of freedom associated with length metric fluctuations generically exhibit nonlocal corrections quadratic in the Weyl curvature beyond the Einstein-Hilbert term. For a two-parameter subclass of area metric Lagrangians, the effective graviton propagator remains free from additional poles, besides the one corresponding to the massless graviton~\cite{Borissova:2023yxs}. Moreover, in contrast to classical gravity, the parity-breaking Barbero-Immirzi (BI) parameter~\cite{BarberoG:1994eia,Immirzi:1996di} enters the classical dynamics of area metrics as a constant which parametrizes a mixing of polarizations for the massless spin-2 mode~\cite{Borissova:2023yxs}. Thereby area metric phenomenology and experimental measurements of the BI parameter may be within reach in the future.

Given that area metrics appear at various instances in quantum gravity and from a physical point of view encode additional degrees of freedom, present above a certain energy scale and turned off in the deep infrared (IR), or low-energy limit, it is essential to understand mathematically under which conditions a generic area metric reduces to an area metric induced by a length metric. The algebraic symmetries of the area metric imply that the area metric field can be decomposed further into irreducible components. These should, however, not be regarded as independent component fields. Rather, in fundamental theories where the area metric defines the spacetime geometry, the area metric field should be massless as an extension of the metric which is massless in Einstein's theory of general relativity. As Einstein's theory of gravity provides a good approximation to the gravitational dynamics in the low-energy limit, we actually observe the metric field as the only surviving massless component of the area metric. There should, therefore, be an intermediate energy scale at which it is reasonable to discuss conditions for the area metric to reduce to a metric. In the effective theory at this intermediate energy scale, we expect that a potential for the area metric exists which gives mass to the non-metric degrees of freedom of the area metric. 
In the second part of this work, we find a robust connection between such a potential and the cosmological constant.

The effective action for the area metric field can be understood as arising from a fundamental theory of quantum gravity. On this basis, in the third part of this work, we investigate the origin of area metrics and area metric actions in the context of string theory. To that end, starting from the Nambu-Goto action defined as the area of the worldsheet of a string, we consider perturbations of the background from a metric background to an area metric background. We will demonstrate that such area metric perturbations introduce a singular interaction in the classically equivalent Polyakov action, whose relevance in the quantum theory remains to be understood.

Our paper is structured as follows. In Section~\ref{Sec:CyclicAreaMetricsGilkey} we first introduce the notion of a cyclic area metric and discuss some of its key properties. In Section~\ref{Sec:ConditionsInduced} we study necessary and sufficient algebraic conditions for an area metric to be induced by a metric. In Section~\ref{Sec:PotentialEnergy} we introduce a potential energy term for the effective area metric action and point out possible connections to low-energy physics and the cosmological constant. In Section~\ref{Sec:WorldSheetAction} we consider area metric perturbations in the context of string theory and derive the resulting modifications to the worldsheet action. We finish with a discussion in Section~\ref{Sec:Discussion}.

\section{Cyclic area metrics}\label{Sec:CyclicAreaMetricsGilkey}

An area metric~\cite{Schuller:2005yt} at a point $p$ on a smooth $d$-dimensional manifold $M$ is a nondegenerate rank-four tensor $G$ with the symmetries
\be\label{eq:GSymmetriesAlgebraic}
G_{\mu\nu\rho\sigma} = - G_{\nu\mu\rho\sigma} = G_{\rho\sigma\mu\nu}\,.
\ee
Nondegeneracy of the area metric is encoded in a nonzero determinant of the area metric matrix $G_{AB}$, where $A,B = 1,\dots, d(d-1)/2$ label antisymmetric index pairs. This allows us to view $G$ as a metric on the space $\Lambda^2 T_p M$ of bivectors and raise bivector indices via the area metric inverse $G^{-1}$, which we denote by $G^{\mu\nu\rho\sigma}$ with upper indices. The area metric inverse is defined by
\be
G_{\mu\nu\rho\sigma}G^{\rho\sigma \alpha\beta} \equiv \delta\indices{_\mu_\nu^\alpha^\beta} \equiv 
\delta_\mu^\alpha \delta_\nu^\beta - \delta_\mu^\beta \delta_\nu^\alpha
\,.
\label{eq:G-inv}
\ee
Here, the generalized delta with four indices defines the identity on the space of area metrics.

In what follows we will impose that the area metric tensor $G$ satisfies in addition to the index exchange symmetries~\eqref{eq:GSymmetriesAlgebraic} also the algebraic Bianchi identity
\be\label{eq:Cyclicity}
G_{\mu[\nu\rho\sigma]} = 0\,.
\ee
In this case, $G$ is called a cyclic area metric. Let us for completeness point out that, in $d=4$ spacetime dimensions, the symmetries~\eqref{eq:GSymmetriesAlgebraic} imply that the cyclicity condition~\eqref{eq:Cyclicity} imposed on an area metric is equivalent to the requirement $G_{\mu\nu\rho\sigma}\tilde{\epsilon}^{\mu\nu\rho\sigma} = 0$, where $\tilde{\epsilon}^{\mu\nu\rho\sigma}$ is the totally antisymmetric Levi-Civita symbol with density weight $+1$. In particular, a cyclic area metric $G$ in $d=4$ has no totally antisymmetric component.

In this paper we will deal with cyclic area metrics throughout. The consideration of cyclic area metrics can be motivated from a mathematical and physical perspective. Mathematically, imposing cyclicity of the area metric allows its components to be unambiguously reconstructed from the norms of bivectors in tangent space without assuming knowledge about dihedral angles between the planes spanned by these bivectors~\cite{Borissova:2022clg}. Such a property of the area metric is analogous to the property of a length metric being fully determined by the lengths of vectors in tangent space, without any knowledge required about angles between these vectors. 

Physically, cyclic area metrics arise in loop quantum gravity and spin foams, as well as in string theory. The semi-classical limit of effective spin foams~\cite{Asante:2020qpa} is described by the Area-Regge action~\cite{Rovelli:1993kc,Barrett:1997tx}. The latter in a lattice continuum limit can be understood as a discretization of an action for a cyclic area metric~\cite{Dittrich:2021kzs,Dittrich:2022yoo}. The same type of actions arises from modified non-chiral Plebanski theories~\cite{DePietri:1998hnx,Alexandrov:2008fs,Speziale:2010cf}, if a only a subset of the simplicity constraints is imposed on the B-field and the other subset is replaced by a potential~\cite{Borissova:2022clg}. Thereby the remaining components of the B-field can be identified with the degrees of freedom of a cyclic area metric~\cite{Borissova:2022clg}. Moreover, the twisted geometry of a coherent simplex has been shown to give rise to a microscopic cyclic area metric associated with such a simplex~\cite{Dittrich:2023ava}. 
In the context of string theory, the infinitesimal area element on the worldsheet can be defined directly through an area metric, and the cyclicity condition arises automatically to remove redundant degrees of freedom in the area metric~\cite{Ho:2015cza}.

The algebraic symmetries~\eqref{eq:GSymmetriesAlgebraic} and algebraic Bianchi identity~\eqref{eq:Cyclicity} turn $G$ into what is called an algebraic curvature tensor. In general spacetime dimension $d$, such a tensor has 
\be
\frac{1}{12}d^2 (d-1)(d+1)
\ee
independent components. For example, in $d=2$ dimensions, a cyclic area metric has only one independent component $G_{0101}$, whereas in $d=3$ dimensions the number $6$ of independent components equals the number of independent components of a length metric. In $d>3$ dimensions, the tensor $G$ encodes more degrees of freedom than a length metric and thus area metric geometry in spacetime dimensions $d=4$ and above can be viewed as a generalization of spacetime geometry based on length metrics.

As an algebraic curvature map, a cyclic area metric admits a Gilkey decomposition~\cite{Gilkey:2001dec,Schuller:2005yt},
\be\label{eq:GilkeyDecomposition}
G= \sum_{I=1}^{N}\sigma_I G_{g^{I}}\,, \quad \sigma_I = \pm 1\,,
\ee
into a sum of area metrics $G_{g^{I}}$ induced by length metrics $g^I$, i.e.,
\be\label{eq:GInduced}
\qty({G_{g^I}})_{\mu\nu\rho\sigma}\equiv g^{I}_{\mu\rho} g^{I}_{\sigma\nu} -g^{I}_{\mu\sigma}g^{I}_{\rho\nu}\,.
\ee
Such a decomposition is not unique and the number $N$ of metrics $g^I$ required to decompose a given cyclic area metric is not known. $N$ has been shown to be bounded from above by~\cite{DiazRamos2004dec,Schuller:2005yt}
\be
N_{\text{max}} = \frac{d(d+1)}{2}\,.
\ee

\section{Conditions for induced area metrics}\label{Sec:ConditionsInduced}
\label{sec:4D-condition}

In this Section, we will investigate necessary and sufficient conditions for a cyclic area metric $G_{\mu\nu\rho\sigma}$ to be induced by a single length metric $g_{\mu\nu}$, as in equation~\eqref{eq:GilkeyDecomposition} with $N=1$. From now on we will drop the adjective ``cyclic" when referring to the area metric $G$.

In odd spacetime dimensions $d = 2n+1$ with  $n\in\mathbb{N}$,
one can construct rank-two tensors from the area metric given by~\cite{Ho:2015cza}
\ba
H^{\mu\nu} &\equiv& \pm \frac{1}{2^n (2n)!} \abs{\det(G)}^{-\frac{1}{2n}} 
\tilde{\epsilon}^{\mu \mu_1 \mu_2 \cdots \mu_{2n}} \tilde{\epsilon}^{\nu \nu_1 \nu_2 \cdots \nu_{2n}}
G_{\mu_1 \mu_2 \nu_1 \nu_2} \cdots G_{\mu_{2n-1} \mu_{2n} \nu_{2n-1} \nu_{2n}} \, ,
\label{eq:H}
\\
H'_{\mu\nu} &\equiv&\pm  \frac{ 1}{2^n (2n)!} \abs{\det(G)}^{\frac{1}{2n}} 
\tilde{\epsilon}_{\mu \mu_1 \mu_2 \cdots \mu_{2n}} \tilde{\epsilon}_{\nu \nu_1 \nu_2 \cdots \nu_{2n}}
G^{\mu_1 \mu_2 \nu_1 \nu_2} \cdots G^{\mu_{2n-1} \mu_{2n} \nu_{2n-1} \nu_{2n}} \, ,
\label{eq:H'}
\ea
where the $G$'s with upper indices denote the inverse area metric defined in equation~\eqref{eq:G-inv}.
Here the ``$\pm$" sign is determined by the sign of $\det(G)$ and $\tilde{\epsilon}$ with upper or lower indices denotes, respectively, the totally antisymmetric Levi-Civita symbol with density weight $\pm 1$ in $d=2n+1$ dimensions.

With the above definitions, the necessary and sufficient condition for an area metric $G$ to be induced by a metric $g$, i.e., that $G =G_g$, is
\ba\label{eq:condition-2}
G_{\mu\nu\rho\sigma} = H'_{\mu\rho} H'_{\nu\sigma} - H'_{\mu\sigma} H'_{\nu\rho}\,,
\qquad \mbox{or equivalently} \qquad
G^{\mu\nu\rho\sigma} = H^{\mu\rho} H^{\nu\sigma} - H^{\mu\sigma} H^{\nu\rho}\,.
\ea
It is easy to show that the above conditions are necessary by verifying that both equations in~\eqref{eq:condition-2} are satisfied when $G =G_g$. In this case,
\be
H'_{\mu\nu} = g_{\mu\nu}
\qquad \mbox{and} \qquad
H^{\mu\nu} = g^{\mu\nu}\,.
\ee
The sufficiency of~\eqref{eq:condition-2} is manifest,
as we can simply define $H'_{\mu\nu}$ to be the metric $g_{\mu\nu}$, or equivalently,  $H^{\mu\nu}$ to be the inverse metric $g^{\mu\nu}$.

Rank-two tensors, such as $H^{\mu\nu}$ in equation~\eqref{eq:H} and $H'_{\mu\nu}$ in equation~\eqref{eq:H'}, do not exist in even spacetime dimensions.
However, if the area metric is induced by a metric, i.e., $G=G_g$,
the dimensionally reduced area metric would also be given by the dimensionally reduced metric.
For instance,
in $d=4$ dimensions,
choosing a direction $a_{\mu}$ for dimensional reduction,
the three-dimensional Levi-Civita symbol can be realized as 
$a_{\mu} \tilde{\epsilon}^{\mu\nu\rho\sigma}$.
Equation~\eqref{eq:H} for $d=3$ dimensions implies
that, with $a_{\mu}$ projected out,
\be\label{eq:Hpropto}
H^{\mu\nu} \propto \frac{1}{4} a_{\rho} \tilde{\epsilon}^{\rho\mu\alpha\beta} a_{\sigma} \tilde{\epsilon}^{\sigma\nu\gamma\delta} G_{\alpha\beta\gamma\delta}\,,
\ee
is expected to be identified with $g^{\mu\nu}$ when $G = G_g$.
We have omitted the overall factor $\det(G)^{-\frac{1}{2}}$,
which we expect to correspond to the determinant of the area metric in $d=3$ dimensions.
Instead of a direct evaluation,
it is easier to retrieve this overall factor by dimensional analysis later.

On the other hand, 
the dimensional reduction of the area metric $G$ leads to
\be\label{eq:Gaa=g}
G^{\mu\rho\nu\sigma} a_{\rho} a_{\sigma} = \frac{1}{4} g^{\mu\nu} \, .
\ee
According to equations~\eqref{eq:Hpropto} and~\eqref{eq:Gaa=g}, if $H^{\mu\nu} = g^{\mu\nu}$ holds for an arbitrary direction $a_{\mu}$ of dimensional reduction, we need
\be
G^{\mu\rho\nu\sigma} \propto \frac{1}{4^2}  \tilde{\epsilon}^{\mu\rho\alpha\beta} \tilde{\epsilon}^{\nu\sigma\gamma\delta}G_{\alpha\beta\gamma\delta}\,.
\ee
The overall factor can be determined by dimensional analysis. Herewith it follows that~\cite{Ho:2015cza}
\be\label{eq:condition-0}
G^{\mu\rho\nu\sigma} = \pm \frac{1}{4^2} \abs{\det(G)}^{-\frac{1}{3}} \tilde{\epsilon}^{\mu\rho\alpha\beta}\tilde{\epsilon}^{\nu\sigma\gamma\delta}G_{\alpha\beta\gamma\delta}\,.
\ee
One can check that equation~\eqref{eq:condition-0} holds whenever the area metric is induced by a metric. Therefore it represents a necessary condition for $G= G_g$.

In general,
for even spacetime dimensions $d = 2n+2 $ with $n\in \mathbb{N} \cup \{0\}$,
the analogous condition to equation~\eqref{eq:condition-0} is~\cite{Ho:2015cza}
\be\label{eq:evenD-cond-1}
G^{\mu\nu\rho\sigma} = 
\pm \frac{ 1}{2^{n+2} (2n)!} \abs{\det(G)}^{-\frac{1}{2n+1}} 
\tilde{\epsilon}^{\mu \nu \mu_1 \mu_2 \cdots \mu_{2n}} \tilde{\epsilon}^{\rho\sigma \nu_1 \nu_2 \cdots \nu_{2n}}
G_{\mu_1 \mu_2 \nu_1 \nu_2} \cdots G_{\mu_{2n-1} \mu_{2n} \nu_{2n-1} \nu_{2n}} \, ,
\ee
where the sign ``$\pm$" is again given by the sign of $\det(G)$.
Similarly, repeating the argument above with $H$ replaced by $H'$,
we find
\be\label{eq:evenD-cond-2}
G_{\mu\nu\rho\sigma} =\pm
\frac{ 1}{2^{n+2} (2n)!} \abs{\det(G)}^{\frac{1}{2n+1}} 
\tilde{\epsilon}_{\mu\nu \mu_1 \mu_2 \cdots \mu_{2n}} \tilde{\epsilon}_{\rho\sigma \nu_1 \nu_2 \cdots \nu_{2n}}
G^{\mu_1 \mu_2 \nu_1 \nu_2} \cdots G^{\mu_{2n-1} \mu_{2n} \nu_{2n-1} \nu_{2n}} \, .
\ee
Both equations~\eqref{eq:evenD-cond-1} and \eqref{eq:evenD-cond-2} are necessary conditions for $G= G_g$.
In the following, we shall see how close they are to also being a sufficient condition.

Considering $d=4$ and multiplying equation~\eqref{eq:condition-0} by the area metric $G$ on both sides, we find its equivalent expression as
\be
\frac{1}{4^2}G_{\mu\nu \alpha\beta} \tilde{\epsilon}^{\alpha\beta\gamma\delta}G_{\gamma\delta \lambda\tau} \tilde{\epsilon}^{\lambda\tau \rho\sigma} =  
\pm 
\frac{1}{2}
\abs{\det(G)}^{\frac{1}{3}}
{\delta}\indices{_\mu_\nu^\rho^\sigma}\,.
\label{eq:condition-1}
\ee
In the context of pre-metric electrodynamics in four-dimensional spacetime, a condition of the form~\eqref{eq:condition-1} without the area metric determinant factor on the right hand side, when imposed on the non-axion part of the constitutive tensor of spacetime (which has the same algebraic symmetries as the area metric $\noteJB{G}$), is known as ``closure relation" or ``electric and magnetic reciprocity"~\cite{Obukhov:1999ug,Hehl:1999wb,Hehl:1999bt,Gross:2001sa,Rubilar:2002vfs}. In particular, in analogy to pre-metric electrodynamics, in both Euclidean and Lorentzian signatures as indicated by the capital letters $E$ and $L$, respectively, one can introduce a duality operator from the area metric which acts on $2$-forms $B_{\mu\nu}$ as
\ba\label{eq:DualityOperator}
^{\#_E } B_{\mu\nu}& \equiv & \frac{1}{4} \abs{\det(G)}^{\frac{1}{6}}
\tilde{\epsilon}_{\mu\nu\rho\sigma} G^{\rho\sigma \alpha\beta}B_{\alpha\beta}\,,\\
^{\#_L } B_{\mu\nu}& \equiv & \frac{\imath}{4} \abs{\det(G)}^{\frac{1}{6}}
\tilde{\epsilon}_{\mu\nu\rho\sigma} G^{\rho\sigma \alpha\beta}B_{\alpha\beta}\,.
\ea
Herewith the condition~\eqref{eq:condition-1} can be stated as the requirement that the duality operator applied twice should up to a sign be equivalent to the identity,
\be\label{eq:DualityOperatorCondition}
\qty(^{\#_{E,L}})^2 = \pm \mathbb{I}\,,
\ee
where the ``$+$" sign holds in Euclidean signature and the ``$-$" sign holds in Lorentzian signature. In turn, if the condition~\eqref{eq:condition-1} is taken into account without the determinant factor, a closure relation imposed on the non-axion part of the constitutive tensor of electrodynamics in four spacetime dimensions, reduces its number of independent components from $20$ down to $9$, see, e.g.,~\cite{Obukhov:1999ug,Hehl:1999wb,Hehl:1999bt,Gross:2001sa,Rubilar:2002vfs} and references therein for details of how such a condition is related to the condition for the nonexistence of birefrigence in vacuum. More concretely, solutions to equation~\eqref{eq:DualityOperatorCondition} in Lorentzian signature and without the determinant factor, have been found by decomposing the matrix $\tilde{G}$ into $3\times 3$ block matrices and investigating the resulting conditions on these block matrices for different cases depending on the regularity of the block matrices along the diagonal~\cite{Obukhov:1999ug,Hehl:1999wb,Hehl:1999bt,Gross:2001sa,Rubilar:2002vfs}. Subsequently, the spacetime metric in the context of electrodynamics can, up to a conformal factor, be extracted by an analysis of the propagation of electromagnetic waves~\cite{Obukhov:1999ug,Hehl:1999wb,Hehl:1999bt,Gross:2001sa,Rubilar:2002vfs}, see also~\cite{Schuller:2005ru}. Such an approach is complementary to the Urbantke formula, which allows us to define a unique effective metric from an $su(2)$-valued $2$-form~\cite{Urbantke:1984eb}.

Let us note that in the literature on pre-metric electrodynamics~\cite{Obukhov:1999ug,Hehl:1999wb,Hehl:1999bt,Gross:2001sa,Rubilar:2002vfs}, a condition of the form~\eqref{eq:condition-1} or~\eqref{eq:DualityOperatorCondition} has been studied as a possible condition, which is both necessary and sufficient for a tensor with the same symmetries as the area metric, to be induced from an ordinary metric $g$. This holds, however, only if we allow the metric $g$ to be complex.
	Notice that the condition \eqref{eq:condition-1} is invariant under the transformation
	\be
	G\,\,\, \rightarrow \,\,\, -G\,.
	\ee
Under this transformation, the area metric determinant transforms by multiplication with a factor $(-1)^{d(d-1)/2}$ where $d=4$. Thus, the ``$\pm$" sign appearing in equation~\eqref{eq:condition-1} and defined as the sign of $\det(G)$, is not affected by the above transformation.
	Now, if an area metric $G$ is induced by a metric $g$, then
	after this transformation,
	the area metric $-G$ is defined through the two possible metrics $\pm i g$. This holds because the formula for an induced area metric, cf.~equation~\eqref{eq:GilkeyDecomposition} with $N=1$, is a quadratic expression in $g$.
	It is thus clear that the condition~\eqref{eq:condition-1} alone
	is not sufficient to guarantee that the area metric is given by a real metric $g$,
	as already anticipated in~\cite{Ho:2015cza}.

Let us for completeness demonstrate perturbatively and in Euclidean signature for simplicity, that the condition~\eqref{eq:condition-1} imposed on the area metric, including the determinant factor, indeed reduces the number $20$ of independent components of a cyclic area metric in $d=4$ dimensions down to $10$, as required for a length metric.

To that end, let us assume an area metric $G_{\mu\nu\rho\sigma}$ which is induced by 
a metric $g_{\mu\nu}$. Since equation~\eqref{eq:condition-1} is invariant under general coordinate transformations,
we can choose $g_{\mu\nu}$ to be a diagonal matrix at any given point in spacetime.
In Euclidean signature, we can choose $g_{\mu\nu}$ to be given locally by the flat Euclidean metric $ \delta_{\mu\nu}$
without loss of generality.  In this perturbative framework, we can raise and lower indices by means of the background metric $\delta_{\mu\nu}$.
With area metric perturbations denoted by $a_{\mu\nu\rho\sigma}$, we write
\be\label{eq:Expansion}
G_{\mu\nu\rho\sigma} = 
\delta_{\mu\rho}\delta_{\sigma\nu} -  \delta_{\mu\sigma}\delta_{\rho\nu} 
+ a_{\mu\nu\rho\sigma}\,.
\ee
The tensor of area metric perturbations $a_{\mu\nu\rho\sigma}$ satisfies the same algebraic symmetries as the cyclic area metric. In particular, it can be decomposed into irreducible representations of the group $SO(4)$ according to
\be\label{eq:Decomposition}
a_{\mu\nu\rho\sigma} \equiv 
\frac{1}{2}
h \delta_{\mu\nu\rho\sigma} +  2\qty(\tilde{h}_{\mu[\rho}\delta_{\sigma]\nu} - \tilde{h}_{\nu[\rho}\delta_{\sigma]\mu}) + \omega^+_{\mu\nu\rho\sigma} + \omega^-_{\mu\nu\rho\sigma} \,\,\,\in \,\,\, (0,0)\oplus (1,1) \oplus (2,0) \oplus (0,2)\,,
\ee
see, e.g.,~\cite{Borissova:2023yxs}. 
The tensor $\delta_{\mu\nu\rho\sigma}$ here is defined as the tensor
$\delta_{\mu\nu}{}^{\alpha\beta}$ introduced~\eqref{eq:G-inv}
and indices lowered with respect to the background metric $\delta_{\rho\alpha}$ and $\delta_{\sigma\beta}$.
The decomposition~\eqref{eq:Decomposition} is analogous to the Ricci-Weyl decomposition of the Riemann tensor. 
The tensor $a_{\mu\nu\rho\sigma}$,
and similarly $\omega^{\pm}_{\mu\nu\rho\sigma}$,
can be treated as matrices by viewing the first two (antisymmetrized) indices as a single index $A = [\mu\nu]$ and the latter two as another one $B = [\rho\sigma]$,
so that we can define the trace of $a$ and $\omega^{\pm}$.
In equation~\eqref{eq:Decomposition},
$h$ is proportional to the trace of $a$, $\tilde{h}$ is symmetric and traceless, i.e., $\tilde{h}_{\mu\nu}\delta^{\mu\nu}=0$, whereas the ``Weyl" components $\omega^\pm$ are fully traceless (i.e., $\delta^{\mu\rho}\omega^\pm_{\mu\nu\rho\sigma} = 0$ and similarly for other pairs of contracted indices), and satisfy the self- and anti-self-duality equations with respect to the background $\delta_{\mu\nu}$,
\be
\frac{1}{2} \epsilon\indices{_\mu_\nu^\alpha^\beta} \omega\indices{^\pm_\alpha_\beta_\rho_\sigma} = \pm \omega\indices{^\pm_\mu_\nu_\rho_\sigma} \,.
\ee
Inserting the expansion~\eqref{eq:Expansion} together with the parametrization~\eqref{eq:Decomposition} into the left hand side of equation~\eqref{eq:condition-1}, to first order in the perturbations $\qty(h,\tilde{h},\omega^+,\omega^-)$ leads to
\be
\frac{1}{4^2} G_{\mu\nu \alpha\beta} \tilde{\epsilon}^{\alpha\beta\gamma\delta} G_{\gamma\delta \lambda\tau} \tilde{\epsilon}^{\lambda\tau \rho\sigma} =
\frac{1}{2}
{\delta}\indices{_\mu_\nu^\rho^\sigma}  + 
\frac{1}{2}
{\delta}\indices{_\mu_\nu^\rho^\sigma}h + \omega\indices{^+ _\mu_\nu ^\rho^\sigma} + \omega\indices{^- _\mu_\nu ^\rho^\sigma} + \mathcal{O}\qty(a\qty(h,\tilde{h},\omega^+,\omega^-)^2)\,.
\ee
On the other hand, regarding the right-hand side of equation~\eqref{eq:condition-1}, to the first order, the area metric determinant is given by~\footnote{This result can be verified using, e.g., the relation $\delta\det(\noteJB{G}) = \det(\noteJB{G}) \Tr(\noteJB{G}^{-1}\delta\noteJB{G})$,
which can be easily derived from the identity $\log(\det(\noteJB{G})) = \Tr(\log\noteJB{G})$
by taking the variation $\delta\noteJB{G}$ on both sides.}
\be
\det(G) = 1 + 3 h + \mathcal{O}\qty(a\qty(h,\tilde{h},\omega^+,\omega^-)^2)\,.
\ee
Thus, on the right-hand side of the condition~\eqref{eq:condition-1}, to first order in $\qty(h,\tilde{h},\omega^+,\omega^-)$, we obtain a contribution of $h$ multiplied by a factor of one. Herewith the terms proportional to the trace $h$ cancel on both sides of equation~\eqref{eq:condition-1}. As a consequence, the condition~\eqref{eq:condition-1} imposed to first order in area metric perturbations around a flat Euclidean background, sets the Weyl components of the area metric to zero, $\omega^+_{\mu\nu\rho\sigma} = \omega^-_{\mu\nu\rho\sigma} \equiv 0$, whereas the $10$ degrees of freedom encoded in the trace mode $h$ and symmetric traceless tensor $\tilde{h}_{\mu\nu}$ are left free. The latter can be associated with symmetric length metric fluctuations $h_{\mu\nu}$ defined by~\cite{Borissova:2023yxs}
\be\label{eq:LengthMetricFluctuations}
h_{\mu\nu} = \tilde{h}_{\mu\nu} + \frac{1}{4}h \delta_{\mu\nu}\,.
\ee

The above statements can be generalized to higher even spacetime dimensions. In appendix~\ref{sec:appendix} we show that equation~\eqref{eq:evenD-cond-1} is a sufficient condition for 
area metrics that are small deformations of an area metric induced by a metric, to reduce to the induced area metric.

\section{Potential energy of the area metric}\label{Sec:PotentialEnergy}

Our universe is observed to be well described by an ordinary metric rather than an area metric. In this section, we will consider a possible potential energy term contained in the low-energy effective gravitational action for the area metric and its relation to the gravitational action for the length metric.
We remind the reader that here by ``low-energy" we have in mind an intermediate energy scale at which a potential for the area metric is expected to give rise to masses for the non-metric degrees of freedom.

First, let us note that a volume element with scalar density $+1$ for a generally covariant area metric action can defined by the area metric determinant as
\be\label{eq:VG}
\abs{\det(G)}^{\frac{1}{6}}\dd[4]{x}\,.
\ee
Different from the volume element built from the determinant of a length metric, the volume element built from the area metric determinant is not the unique scalar invariant with density weight $+1$ available to define the constant part of a generally covariant area metric Lagrangian~\cite{Ho:2015cza}. For an area metric, there exist many possible invariants with density weight $+1$ which can enter the constant part of a generally covariant action and contribute to a notion of the potential energy of the area metric. Our ansatz below is general, as we do not specify the potential function $V_{G}$. This function can, for instance, contain a linear combination of other density $+1$ invariants, each in turn multiplied by a factor of $\abs{\det(G)}^{-\frac{1}{6}}$ with possibly independent coupling constants.
	
Thus, we consider the potential energy term in the area metric action of the general form
\be\label{eq:potential}
\int \dd[4]{x} \abs{\det(G)}^{\frac{1}{6}} \, V_{G}\qty(G) \, .
\ee
As part of the action, the potential density $V_{G}$ must behave as a scalar.
In particular, it needs to be invariant under a rescaling of the coordinates,
\ba
x^{\mu} &\rightarrow& \lambda x^{\mu} \, ,
\\
G_{\mu\nu\rho\sigma} &\rightarrow& \lambda^{-4} G_{\mu\nu\rho\sigma} \,.
\ea
Since the measure $\int \dd[4]{x} \abs{\det(\noteJB{G})}^{\frac{1}{6}}$ 
is invariant under this scaling transformation,
the invariance of the action~\eqref{eq:potential} requires that $V_{\noteJB{G}}$ be invariant.
Under an infinitesimal scaling transformation $\lambda = 1 + \epsilon$,
the variation of $\noteJB{G}$ is
\be
\delta G_{\mu\nu\rho\sigma} = - 4 \epsilon G_{\mu\nu\rho\sigma}\,,
\ee
and therefore the corresponding generator on the space of functions of $G$ is given by
\footnote{
Here the derivatives of $G$ by definition 
treat all components of $G$ as independent variables.
}
\be
\delta_{G} \equiv \delta G_{\mu\nu\rho\sigma}
\frac{\partial}{\partial G_{\mu\nu\rho\sigma}}
= - 4 G_{\mu\nu\rho\sigma} \frac{\partial}{\partial G_{\mu\nu\rho\sigma}}\,.
\ee
Therefore,
assuming general covariance
(so that $V_{G}$ depends on $x$ only through $G$),
the invariance of $V_{G}$ states that
\be\label{eq:GdVdG=0}
\delta_{G}V_{G}\qty(G) \equiv -4
G_{\mu\nu\rho\sigma} \frac{\partial V_{G}\qty(G)}{\partial G_{\mu\nu\rho\sigma}} = 0\,.
\ee

For a low-energy configuration, one may ignore the kinetic term when applying the least action principle. Varying
the potential term~\eqref{eq:potential} leads to the condition
\be
\frac{\partial V_{G}\qty(G)}{\partial G_{\mu\nu\rho\sigma}}
+ \frac{1}{6} V_{G}\qty(G)G^{\mu\nu\rho\sigma}
= 0 \, .
\ee
Contracting the indices in this equation with $G_{\mu\nu\rho\sigma}$
and using equation~\eqref{eq:GdVdG=0},
we find
\be
V_{G}\qty(G) = 0
\ee
as a result of 
the least action principle for low-energy configurations. In other words, the potential energy of the area metric defined in equation~\eqref{eq:potential} is always zero onshell of the potential action.

As a simple example for the potential $V_{G}\qty(G)$,
the result of the previous section suggests to consider
\be
V_{G}\qty(G) = \frac{\gamma}{2} \,
G^{\mu\nu\alpha\beta}G_{\rho\sigma\gamma\delta} E\indices{_\mu_\nu^\rho^\sigma}E\indices{_\alpha_\beta^\gamma^\delta}\,,
\ee
where
\be
E\indices{_\mu_\nu^\rho^\sigma} \equiv
\frac{1}{4^2} \abs{\det(G)}^{- \frac{1}{3}} G_{\mu\nu \alpha\beta} \tilde{\epsilon}^{\alpha\beta\gamma\delta} G_{\gamma\delta \lambda\tau} \tilde{\epsilon}^{\lambda\tau \rho\sigma} \mp 
\frac{1}{2}
{\delta}\indices{_\mu_\nu^\rho^\sigma}\,.
\ee
The above potential evaluates to zero when the area metric is iduced by a metric and thus $E=0$ holds. Moreover, generically, at second order in the perturbations around an induced area metric, such a potential would lead to mass terms for the (non-metric) Weyl components $\omega^+$ and $\omega^-$ of the area metric~\eqref{eq:Decomposition}, and in this case their masses would be determined by the coupling constant $\gamma$.

Note that the argument above, for the value of $V_{G}$ at its minimum to be zero, is rather general. One can straightforwardly extend this argument to area metrics in higher dimensions and to higher-volume metrics.~\footnote{Note that in $d=4$ dimensions, a three-volume metric and higher-volume metrics define a coarser geometry than area metric geometry~\cite{Schuller:2009hn}.}
One may also wonder what happens if we repeat the same argument for a length metric $g$, starting from a Lagrangian $\sqrt{\abs{\det(g)}}V_{g}(g)$. It appears that we would similarly conclude that $V_g(g)=0$ onshell of the non-derivative action. However, general covariance forbids $V_{g}$ to be minimized at any particular metric, so $V_{g}$ can only be the cosmological constant.~\footnote{Thus any generalized background structure from which we can derive the conclusions below must necessarily break an analog of general covariance.}

Observationally, we know that $V_g =\Lambda$ is non-zero, but extremely small. From a quantum-field theoretic perspective on {\it metric gravity}, we do not have a good explanation for the value of the cosmological constant. However, from the point of view of classical {\it area metric gravity},
we note that the potential term~\eqref{eq:potential} for the area metric in the extreme low-energy infrared (IR) limit, when the area metric $G$ reduces to a metric $g$, must reduce to the cosmological constant term in the generally covariant Lagrangian for the metric, i.e.,
\be 
\int \dd[4]{x} \abs{\det(G)}^{\frac{1}{6}} \,  V_{G}\qty(G) \quad\quad  \xrightarrow[G =G_g]{ \text{IR}}\quad \quad \Lambda \int \dd[4]{x} \sqrt{\abs{\det(g)}}\,.
\ee
In this sense, the fact that $V_{G}=0$ is the minimum of $V_{G}$ onshell of the potential action for the area metric and the observation that the universe is well approximated by a metric instead of a general area metric, are correlated with the smallness of the cosmological constant. 

We emphasize that we are not ambitioning to solve the cosmological constant problem, but simply pointing out the connection between the smallness of the cosmological constant and the phenomenological fact that the universe is well approximated by a metric, assuming a fundamental theory for the area metric. That is, the force that drives the area metric towards one induced by a metric is simultaneously also driving the cosmological constant towards zero. 

It should be noted how minimalistic are the arguments that lead to the above conclusions. More generally, all that is needed is the following:
\begin{itemize}
\item[I.] A more general geometric background structure, such as the area metric.
\item[II.] The expectation that this general geometric structure reduces to the metric at low energies.
\item[III.] The assumption that this reduction is realized by extremizing the potential energy of the generalized background, rather than dynamically.
\end{itemize}

We have so far ignored the kinetic terms of the area metric and possible couplings to matter fields. In a dynamical process, the potential energy does not stay at its minimum. It will be interesting to take into account the dynamics of the area metric, for example in a perturbative framework~\cite{Borissova:2023yxs}, and to extract predictions from area metric gravity on cosmology and black hole physics.

\section{Area metric perturbations and the worldsheet action}\label{Sec:WorldSheetAction}

In the previous section, we investigated the effective action for the area metric at the intermediate scale, at which the constraint~\eqref{eq:condition-1} for reduction of the area metric to an ordinary metric is not yet imposed.
In what follows, we will provide an explanation for the possible origin of the effective area metric action in string theory, viewed as a fundamental theory of spacetime geometry.

Let us consider a two-dimensional surface $\Sigma$ as the string worldsheet embedded in a $d$-dimensional spacetime with embedding coordinates $X^\mu (\xi)$, where the $\mu=0,\dots,d-1$ denote spacetime indices and $\xi^i = (\xi^0,\xi^1)$ are the worldsheet coordinates.
If the spacetime is equipped with a metric $g_{\mu\nu}$, the induced metric $h_{ij}$ on $\Sigma$ is defined as
\be\label{eq:MetricInduced}
h_{ij}(\xi) \equiv g_{\mu\nu}(X)\partial_{i}X^\mu (\xi)\partial_j X^\nu (\xi)\,.
\ee
The area of $\Sigma$ defines the Nambu-Goto action,
\be\label{eq:SNambuGoto}
S_{\text{NG}} \equiv {\text{Area}}(\Sigma) = \int_\Sigma \dd{a} = \int_\Sigma \dd[2]\xi \sqrt{\abs{\det(h)}}\,.
\ee
Here $\dd{a}$ denotes the infinitesimal area element on $\Sigma$. The Nambu-Goto action can be obtained from the Polyakov action
\be\label{eq:SPolyakov}
S_{\text{P}} \equiv \frac{1}{2}\int_\Sigma \dd[2]{\xi} \sqrt{\abs{\det(\gamma)}} \gamma^{ij} h_{ij}\,,
\ee
where $\gamma_{ij}$, with $i,j = 0,1$, is the metric on the worldsheet. To that end, the solution to the equation of motion for $\gamma_{ij}$,
\be
\frac{1}{2}\sqrt{\abs{\det(\gamma)}}\gamma^{ij}h_{ij} =  \sqrt{\abs{\det(h)}}\,,
\ee
has to be inserted back into the Polyakov action~\eqref{eq:SPolyakov}.

Using equation~\eqref{eq:MetricInduced}, the determinant of the two-dimensional induced metric $h_{ij}$ can be written as
\be\label{eq:DetH}
\det(h)=\frac{1}{2!}\tilde{\epsilon}^{i j} \tilde{\epsilon}^{k l}h_{i k} h_{j l} = \frac{1}{2} \tilde{\epsilon}^{i j} \tilde{\epsilon}^{k l} \partial_i X^\mu  \partial_j X^\nu  \partial_k X^\rho  \partial_l X^\sigma \qty(G_g)_{\mu\nu\rho\sigma}\,,
\ee
where $G_g$ is the $d$-dimensional area metric defined by the spacetime metric $g$.
Replacing $G_g$ by a generic area metric $G$ gives a generalized Nambu-Goto action defined on an area metric background. From such a perspective, it is natural to consider area metric geometry in string theory, with the cyclicity condition~\eqref{eq:Cyclicity} imposed to remove redundant degrees of freedom in the area metric~\cite{Ho:2015cza}.

Let us now consider the result of perturbing the spacetime geometry from a length metric geometry to an area metric geometry. We will denote fluctuations of the area metric away from a background configuration induced by a length metric $g$, by $a$. Consequently, we replace $G_g$ in equation~\eqref{eq:DetH} by
\be
G
\equiv G_g + a\,.
\ee
As a result 
of the replacement $G_g \mapsto G$, 
the determinant of the induced metric is changed by
\be\label{eq:DetChange}
\det(h) \quad \mapsto \quad \det(h) +  \Delta A\,,
\ee
where
\be
\Delta A \equiv  \frac{1}{2} \tilde{\epsilon}^{i j} \tilde{\epsilon}^{k l} \partial_i X^\mu  \partial_j X^\nu  \partial_k X^\rho  \partial_l X^\sigma a_{\mu\nu\rho\sigma}\,.
\ee
Herewith the action in equation \eqref{eq:SNambuGoto} becomes
	\ba\label{eq:SNG-a}
	S_{\text{NG}} &=& \int_{\Sigma} \dd[2]\xi \sqrt{|\det(h) + \Delta A|} \, .
	\ea
In general,
	an action of the form
	\be\label{eq:SNG-F}
	S_{\text{NG}} = 
	\int_{\Sigma} \dd[2]\xi \; {\cal F}\qty(\det(h))
	\ee
	for an arbitrary function ${\cal F}$,
	is equivalent to the action~\footnote{Note that the 2D metric $\gamma_{ij}$ has more degrees of freedom than the 2D area metric, as we mentioned previously, so it is not a ``generalization'' to replace the worldsheet metric by an area metric.}
	\be\label{eq:SP-F}
	S_{\text{P}} = 
	\int_\Sigma \dd[2]\xi \;
	{\cal F}\qty(\abs{\det(\gamma)}\qty(\frac{1}{2} \gamma^{kl}h_{kl})^2)\,.
	\ee
To show the equivalence,
	note that the equation of motion for $\gamma_{ij}$ derived from \eqref{eq:SP-F} is 
	\be
	h_{ij} = \frac{(\gamma^{kl}h_{kl})}{2} \, \gamma_{ij}\,.
	\ee
	Plugging this result back into the action \eqref{eq:SP-F} reproduces the generalized Nambu-Goto action \eqref{eq:SNG-F}.
	For the action \eqref{eq:SNG-a} of interest, the function $\mathcal{F}$ is defined by
	${\cal F}(x) = \sqrt{x + \Delta A}$.
	It is equivalent to the generalized Polyakov action
	\ba\label{eq:SP-a}
	S_{\text{P}} 
	&=& 
	\int_\Sigma \dd[2]\xi 
	\qty[\abs{\det(\gamma)} \left(\frac{1}{2} \gamma^{kl}h_{kl}\right)^2 + \Delta A]^{1/2}\,.
	\ea
In the conformal gauge, $\gamma_{ij} = e^{\phi(\xi)} \eta_{ij}$,
	the Polyakov action \eqref{eq:SP-a} becomes
	\ba
	S_{\text{P}} &=&
	\int_\Sigma \dd[2]\xi 
	\qty[\qty(\frac{1}{2} \eta^{kl}h_{kl})^2 + \Delta A]^{1/2}
	\\
	&=&
	\frac{1}{2} \int_\Sigma \dd[2]\xi \qty[
	g_{\mu\nu} \partial^i X^{\mu} \partial_i X^{\nu}
	+ \frac{\tilde{\epsilon}^{i j} \tilde{\epsilon}^{k l} \partial_i X^\mu  \partial_j X^\nu  \partial_k X^\rho \partial_l X^\sigma a_{\mu\nu\rho\sigma}}{g_{\alpha\beta} \partial^i X^{\alpha} \partial_i X^{\beta}}
	+ \cdots]\,.
	\ea
	We see that the area metric deformation corresponds to a singular vertex operator
	\be\label{eq:Vertex}
	a_{\mu\nu\rho\sigma} \frac{\partial X^{\mu} \bar{\partial} X^{\nu}  \partial X^{\rho} \bar{\partial} X^{\sigma}}{\partial X^{\alpha} \bar{\partial} X_{\alpha}} \, e^{ik_{\beta} X^{\beta}} + \cdots
	\ee
	with momentum $k$ in the string worldsheet theory for the flat spacetime background.
	This vertex operator transforms properly in the classical theory, but it is not a primary operator in the quantum theory, so we need to add correction terms to make it primary.
	Starting with a sequence of vertex operators 
	\be
	a_{\mu\nu\rho\sigma} \, \partial X^{\mu} \bar{\partial} X^{\nu}  \partial X^{\rho} \bar{\partial} X^{\sigma}\qty(\partial X^{\alpha} \bar{\partial} X_{\alpha})^n \, e^{ik_{\beta} X^{\beta}} + \cdots
	\ee
	with the onshell condition $k^2 = 2(n+1)/\alpha'$ for $n>0$, the vertex operator \eqref{eq:Vertex} is an analytic continuation of the number $n$ to $n = -1$. Since $k^2 = 0$ for $n = -1$, the area metric is a massless field. It will be interesting to see if such a primary operator can be consistently defined in string theory. We leave this problem for future study.

\section{Discussion}\label{Sec:Discussion}

Area metrics play a key role in main candidate approaches to quantum gravity, such as in string theory and loop quantum gravity. Thereby they provide a unified framework for quantum gravity and may provide a ground for establishing connections between these seemingly distinct theories.  At instances where area metrics appear in quantum gravity, they encode physical degrees of freedom beyond those of an ordinary metric. These degrees of freedom must be suppressed by a potential at sufficiently high energy scales
to ensure a consistent limit in which our semiclassical picture of standard-model matter in a background described by Einstein's general relativity is recovered.

On this basis, in this work, we first studied algebraic conditions for a general area metric to reduce to an area metric induced by a metric. Physically, we understand these conditions as realizing the low-energy limit in which the non-metric degrees of freedom originating from an intrinsically quantum theory of gravity are switched off. In loop quantum gravity and spin foams, such a limit can be associated with the partially second-class quantum algebra of constraints reducing to its classical first-class analog by taking the BI constant, which parametrizes the quantum anomaly, to zero~\cite{Dittrich:2008ar,Dittrich:2010ey,Dittrich:2012rj,Asante:2020qpa}. It will be interesting to study the condition provided here for an area metric to reduce to a length metric in the continuum, in the context of area metrics appearing in discrete settings~\cite{Dittrich:2021kzs,Dittrich:2022yoo,Dittrich:2023ava,Asante:2024rrd}, and to establish a relation to area-length or shape-matching constraints for twisted geometries~\cite{Dittrich:2008va,Dittrich:2008ar,Dittrich:2010ey, Dittrich:2012rj,Asante:2024rrd}. 

In the second part of this work, motivated by the fact that our classical universe is well approximated by Einstein gravity for a length metric, we considered the potential energy term in the action for an area metric, which must lead to the frozen dynamics of the non-metric degrees of freedom at low energies. This term is expected to appear in the low-energy effective theory of a fundamental theory, such as string theory in generic area-metric backgrounds or loop quantum gravity and spin foams in the semiclassical regime. At a much lower energy scale, where the description of the spacetime is well approximated by the metric, we assumed this effective potential term to be approximately minimized. Thus, the intermediate energy scale, at which the non-metric degrees of freedom of the area metric 
are massive propagating fields, must lie between the string scale $1/\ell_s$ and the energy scale of the Einstein gravity. 
From the point of view of loop quantum gravity and spin foams, this energy scale is expected to be determined by the value of the BI parameter and the Planck mass~\cite{Borissova:2022clg,Borissova:2023yxs}.

In the third part of this work, we considered the possible origin of the effective area metric action in the context of string theory. In terms of the Nambu-Goto action and the Schild action,
only the area metric is needed, and it is very natural to replace the metric with the area metric,
at least at the classical level. As a result of replacing the metric with an area metric in a perturbative framework around a background induced by a metric, the woldsheet action for a string acquires a correction term which is characterized by a singular vertex operator built from the area metric perturbation. The singular factor $1/\qty(\partial X^{\mu} \bar\partial X_{\mu})$ in the obtained vertex operator suggests a large effect in the IR limit.
At the quantum level,
it is not yet clear whether the area metric deformation of the Polyakov action in the conformal gauge is well defined.
If it is, this would suggest the existence of a new class of (seemingly singular) vertex operators and corresponding backgrounds that can be introduced in string theory, and would significantly modify our understanding of the moduli space, the ``landscape'' of string theory.
On the other hand, the insistence on formulating string worldsheet theory in the background of Minkowski space is a technical convenience rather than a necessity, and one may try to formulate string theory directly in a constant area metric background that is not the deformation of a metric background.

\section*{Acknowledgements}

We thank Bianca Dittrich, Hikaru Kawai and Henry Liao for valuable discussions. J.B.~is supported by an NSERC grant awarded to Bianca Dittrich and a doctoral scholarship by the German Academic Scholarship Foundation.
P.M.H.\ is supported in part by the Ministry of Science and Technology, R.O.C.
(MOST 110-2112-M-002 -016 -MY3),
and by National Taiwan University. Research at Perimeter Institute is supported in part by the Government of Canada through the Department of Innovation, Science and Economic Development Canada and by the Province of Ontario through the Ministry of Colleges and Universities.

%\appendix
\section*{Appendix}
\label{sec:appendix}

In this appendix, we prove that when the area metric is a small deformation of an induced area metric, 
equation~\eqref{eq:evenD-cond-1} is sufficient to imply that the area metric reduces to the one induced by a certain length metric.

Similar to the proof for the $d=4$ case in Section~\ref{sec:4D-condition}, we consider the Euclidean case for simplicity and the area metric expanded in the form~\eqref{eq:Expansion} after choosing a suitable coordinate system. It is straightforward to check that equation~\eqref{eq:evenD-cond-1} is, at the leading order in the area metric deformations $a_{\mu\nu\rho\sigma}$, equivalent to
\be
a_{\mu\nu\rho\sigma} = \frac{1}{2n} \left[
a\indices{_\nu_\lambda_\sigma^\lambda}\delta_{\mu\rho} - a\indices{_\nu_\lambda_\rho^\lambda}\delta_{\mu\sigma}
- a\indices{_\mu_\lambda_\sigma^\lambda}\delta_{\nu\rho} + a\indices{_\mu_\lambda_\rho^\lambda}\delta_{\nu\sigma}
\right]
- \frac{1}{2n(2n+1)} \left(\delta_{\mu\rho}\delta_{\nu\sigma} - \delta_{\mu\sigma}\delta_{\nu\rho}\right) a\indices{_\alpha_\beta^\alpha^\beta}\,,
\ee
where $d=2n +2$ for $n\in \mathbb{N}$.
This allows us to define 
\be
h_{\mu\nu} =
\frac{1}{2n} a\indices{_\mu_\lambda_\nu^\lambda} - \frac{1}{4n(2n+1)} a\indices{_\alpha_\beta^\alpha^\beta} \delta_{\mu\nu}\,.
\ee
As a result,
to the 1st order of the perturbative expansion in $a_{\mu\nu\rho\sigma}$, it holds that
\be
G =G_g\,,
\ee
where
\be
g_{\mu\nu} = \delta_{\mu\nu} + h_{\mu\nu}.
\ee
This completes the proof.

\bibliographystyle{jhep}
\bibliography{references}

\providecommand{\href}[2]{#2}\begingroup\raggedright\begin{thebibliography}{10}

\bibitem{Schuller:2005yt}
F.~P. Schuller and M.~N.~R. Wohlfarth, {\it {Geometry of manifolds with area
  metric: multi-metric backgrounds}},  {\em Nucl. Phys. B} {\bf 747} (2006)
  398--422, [\href{http://arxiv.org/abs/hep-th/0508170}{{\tt hep-th/0508170}}].

\bibitem{Schuller:2005ru}
F.~P. Schuller and M.~N.~R. Wohlfarth, {\it {Canonical differential geometry of
  string backgrounds}},  {\em JHEP} {\bf 02} (2006) 059,
  [\href{http://arxiv.org/abs/hep-th/0511157}{{\tt hep-th/0511157}}].

\bibitem{Punzi:2006hy}
R.~Punzi, F.~P. Schuller, and M.~N.~R. Wohlfarth, {\it {Geometry for the
  accelerating universe}},  {\em Phys. Rev. D} {\bf 76} (2007) 101501,
  [\href{http://arxiv.org/abs/hep-th/0612133}{{\tt hep-th/0612133}}].

\bibitem{Punzi:2006nx}
R.~Punzi, F.~P. Schuller, and M.~N.~R. Wohlfarth, {\it {Area metric gravity and
  accelerating cosmology}},  {\em JHEP} {\bf 02} (2007) 030,
  [\href{http://arxiv.org/abs/hep-th/0612141}{{\tt hep-th/0612141}}].

\bibitem{Ho:2015cza}
P.-M. Ho and T.~Inami, {\it {Geometry of Area Without Length}},  {\em PTEP}
  {\bf 2016} (2016), no.~1 013B03, [\href{http://arxiv.org/abs/1508.05569}{{\tt
  arXiv:1508.05569}}].

\bibitem{Borissova:2023yxs}
J.~N. Borissova, B.~Dittrich, and K.~Krasnov, {\it {Area-metric gravity
  revisited}},  {\em Phys. Rev. D} {\bf 109} (2024), no.~12 124035,
  [\href{http://arxiv.org/abs/2312.13935}{{\tt arXiv:2312.13935}}].

\bibitem{Ryu:2006bv}
S.~Ryu and T.~Takayanagi, {\it {Holographic derivation of entanglement entropy
  from AdS/CFT}},  {\em Phys. Rev. Lett.} {\bf 96} (2006) 181602,
  [\href{http://arxiv.org/abs/hep-th/0603001}{{\tt hep-th/0603001}}].

\bibitem{Swingle:2017blx}
B.~Swingle, {\it {Spacetime from Entanglement}},  {\em Ann. Rev. Condensed
  Matter Phys.} {\bf 9} (2018) 345--358.

\bibitem{Rovelli:1997yv}
C.~Rovelli, {\it {Loop quantum gravity}},  {\em Living Rev. Rel.} {\bf 1}
  (1998) 1, [\href{http://arxiv.org/abs/gr-qc/9710008}{{\tt gr-qc/9710008}}].

\bibitem{Perez:2012wv}
A.~Perez, {\it {The Spin Foam Approach to Quantum Gravity}},  {\em Living Rev.
  Rel.} {\bf 16} (2013) 3, [\href{http://arxiv.org/abs/1205.2019}{{\tt
  arXiv:1205.2019}}].

\bibitem{Dittrich:2023ava}
B.~Dittrich and J.~Padua-Arg\"uelles, {\it {Twisted geometries are area-metric
  geometries}},  {\em Phys. Rev. D} {\bf 109} (2024), no.~2 026002,
  [\href{http://arxiv.org/abs/2302.11586}{{\tt arXiv:2302.11586}}].

\bibitem{Rovelli:1993kc}
C.~Rovelli, {\it {The Basis of the Ponzano-Regge-Turaev-Viro-Ooguri quantum
  gravity model in the loop representation basis}},  {\em Phys. Rev. D} {\bf
  48} (1993) 2702--2707, [\href{http://arxiv.org/abs/hep-th/9304164}{{\tt
  hep-th/9304164}}].

\bibitem{Barrett:1997tx}
J.~W. Barrett, M.~Rocek, and R.~M. Williams, {\it {A Note on area variables in
  Regge calculus}},  {\em Class. Quant. Grav.} {\bf 16} (1999) 1373--1376,
  [\href{http://arxiv.org/abs/gr-qc/9710056}{{\tt gr-qc/9710056}}].

\bibitem{Asante:2020qpa}
S.~K. Asante, B.~Dittrich, and H.~M. Haggard, {\it {Effective Spin Foam Models
  for Four-Dimensional Quantum Gravity}},  {\em Phys. Rev. Lett.} {\bf 125}
  (2020), no.~23 231301, [\href{http://arxiv.org/abs/2004.07013}{{\tt
  arXiv:2004.07013}}].

\bibitem{Dittrich:2021kzs}
B.~Dittrich, {\it {Modified Graviton Dynamics From Spin Foams: The Area Regge
  Action}},  \href{http://arxiv.org/abs/2105.10808}{{\tt arXiv:2105.10808}}.

\bibitem{Dittrich:2022yoo}
B.~Dittrich and A.~Kogios, {\it {From spin foams to area metric dynamics to
  gravitons}},  {\em Class. Quant. Grav.} {\bf 40} (2023), no.~9 095011,
  [\href{http://arxiv.org/abs/2203.02409}{{\tt arXiv:2203.02409}}].

\bibitem{Dittrich:2008ar}
B.~Dittrich and J.~P. Ryan, {\it {Phase space descriptions for simplicial 4d
  geometries}},  {\em Class. Quant. Grav.} {\bf 28} (2011) 065006,
  [\href{http://arxiv.org/abs/0807.2806}{{\tt arXiv:0807.2806}}].

\bibitem{Dittrich:2010ey}
B.~Dittrich and J.~P. Ryan, {\it {Simplicity in simplicial phase space}},  {\em
  Phys. Rev. D} {\bf 82} (2010) 064026,
  [\href{http://arxiv.org/abs/1006.4295}{{\tt arXiv:1006.4295}}].

\bibitem{Dittrich:2012rj}
B.~Dittrich and J.~P. Ryan, {\it {On the role of the Barbero-Immirzi parameter
  in discrete quantum gravity}},  {\em Class. Quant. Grav.} {\bf 30} (2013)
  095015, [\href{http://arxiv.org/abs/1209.4892}{{\tt arXiv:1209.4892}}].

\bibitem{Borissova:2022clg}
J.~N. Borissova and B.~Dittrich, {\it {Towards effective actions for the
  continuum limit of spin foams}},  {\em Class. Quant. Grav.} {\bf 40} (2023),
  no.~10 105006, [\href{http://arxiv.org/abs/2207.03307}{{\tt
  arXiv:2207.03307}}].

\bibitem{DePietri:1998hnx}
R.~De~Pietri and L.~Freidel, {\it {so(4) Plebanski action and relativistic spin
  foam model}},  {\em Class. Quant. Grav.} {\bf 16} (1999) 2187--2196,
  [\href{http://arxiv.org/abs/gr-qc/9804071}{{\tt gr-qc/9804071}}].

\bibitem{Alexandrov:2008fs}
S.~Alexandrov and K.~Krasnov, {\it {Hamiltonian Analysis of non-chiral
  Plebanski Theory and its Generalizations}},  {\em Class. Quant. Grav.} {\bf
  26} (2009) 055005, [\href{http://arxiv.org/abs/0809.4763}{{\tt
  arXiv:0809.4763}}].

\bibitem{Speziale:2010cf}
S.~Speziale, {\it {Bi-metric theory of gravity from the non-chiral Plebanski
  action}},  {\em Phys. Rev. D} {\bf 82} (2010) 064003,
  [\href{http://arxiv.org/abs/1003.4701}{{\tt arXiv:1003.4701}}].

\bibitem{BarberoG:1994eia}
J.~F. Barbero~G., {\it {Real Ashtekar variables for Lorentzian signature space
  times}},  {\em Phys. Rev. D} {\bf 51} (1995) 5507--5510,
  [\href{http://arxiv.org/abs/gr-qc/9410014}{{\tt gr-qc/9410014}}].

\bibitem{Immirzi:1996di}
G.~Immirzi, {\it {Real and complex connections for canonical gravity}},  {\em
  Class. Quant. Grav.} {\bf 14} (1997) L177--L181,
  [\href{http://arxiv.org/abs/gr-qc/9612030}{{\tt gr-qc/9612030}}].

\bibitem{Gilkey:2001dec}
P.~B. Gilkey, {\it {Geometric properties of natural operators defined by the
  Riemann curvature tensor}},  {\em World Scientific)} (2001).

\bibitem{DiazRamos2004dec}
J.~Diaz-Ramos, B.~Fiedler, E.~Garcia-Rio, and P.~B. Gilkey {\em J. Geom. Meth.
  Mod. Phys. 1} {\bf 711} (2004).

\bibitem{Obukhov:1999ug}
Y.~N. Obukhov and F.~W. Hehl, {\it {Space-time metric from linear
  electrodynamics}},  {\em Phys. Lett. B} {\bf 458} (1999) 466--470,
  [\href{http://arxiv.org/abs/gr-qc/9904067}{{\tt gr-qc/9904067}}].

\bibitem{Hehl:1999wb}
F.~W. Hehl, Y.~N. Obukhov, and G.~F. Rubilar, {\it {Space-time metric from
  linear electrodynamics. 2.}},  in {\em {International European Conference on
  Gravitation: Journees Relativistes 99}}, 9, 1999.
\newblock \href{http://arxiv.org/abs/gr-qc/9911096}{{\tt gr-qc/9911096}}.

\bibitem{Hehl:1999bt}
F.~W. Hehl and Y.~N. Obukhov, {\it {How does the electromagnetic field couple
  to gravity, in particular to metric, nonmetricity, torsion, and curvature?}},
   {\em Lect. Notes Phys.} {\bf 562} (2001) 479,
  [\href{http://arxiv.org/abs/gr-qc/0001010}{{\tt gr-qc/0001010}}].

\bibitem{Gross:2001sa}
A.~Gross and G.~F. Rubilar, {\it {On the derivation of the space-time metric
  from linear electrodynamics}},  {\em Phys. Lett. A} {\bf 285} (2001)
  267--272, [\href{http://arxiv.org/abs/gr-qc/0103016}{{\tt gr-qc/0103016}}].

\bibitem{Rubilar:2002vfs}
G.~F. Rubilar, {\it {Linear pre-metric electrodynamics and deduction of the
  light cone}},  {\em Annalen Phys.} {\bf 11} (2002) 717--782,
  [\href{http://arxiv.org/abs/0706.2193}{{\tt arXiv:0706.2193}}].

\bibitem{Urbantke:1984eb}
H.~Urbantke, {\it {ON INTEGRABILITY PROPERTIES OF SU(2) YANG-MILLS FIELDS. I.
  INFINITESIMAL PART}},  {\em J. Math. Phys.} {\bf 25} (1984), no.~7
  2321--2324.

\bibitem{Schuller:2009hn}
F.~P. Schuller, C.~Witte, and M.~N.~R. Wohlfarth, {\it {Causal structure and
  algebraic classification of area metric spacetimes in four dimensions}},
  {\em Annals Phys.} {\bf 325} (2010) 1853--1883,
  [\href{http://arxiv.org/abs/0908.1016}{{\tt arXiv:0908.1016}}].

\bibitem{Asante:2024rrd}
S.~K. Asante and T.~Brysiewicz, {\it {Solving the area-length systems in
  discrete gravity using homotopy continuation}},
  \href{http://arxiv.org/abs/2402.17080}{{\tt arXiv:2402.17080}}.

\bibitem{Dittrich:2008va}
B.~Dittrich and S.~Speziale, {\it {Area-angle variables for general
  relativity}},  {\em New J. Phys.} {\bf 10} (2008) 083006,
  [\href{http://arxiv.org/abs/0802.0864}{{\tt arXiv:0802.0864}}].

\end{thebibliography}\endgroup

\end{document}